\documentstyle[10pt]{article}

\input epsf.sty
\def\etal{{\it et al\ }} 
\def\Mpc{$h^{-1}$~{\rm Mpc}}
\def\apj{{\it Astrophys. J.}}
\def\mn{{\it Mon. Not. R. astr. Soc.}}
\def\aa{{\it Astron. Astrophys.}}
\begin{document}

\title{THE OSCILLATING CLUSTER CORRELATION FUNCTION}
\author{JAAN EINASTO\\
Tartu Observatory, EE-2444, Estonia}  

\maketitle

\section{Introduction}

According to the traditional view galaxies are clustered on small scales, and
on large scales the distribution of galaxies and clusters of galaxies is
random (i.e. scale-free). Such a distribution is expected for the currently
popular structure formation scenario based on the evolving dynamics of the
dark matter during the history of the Universe.  Quantitatively this behaviour
is described by the correlation function of galaxies and clusters of galaxies
which has a high peak at small separations but approaches zero at about 30
\Mpc\ for galaxies and at about 70 \Mpc\ for clusters of galaxies (in this
paper we express the Hubble constant as $H_0=h~100~{\rm km~s^{-1}~Mpc}^{-1}$).
Initially these scales were thought to be the scales of the transition to
homogeneity. However, in the distribution of galaxies voids of diameter up to
50 \Mpc\ were discovered, voids in the distribution of clusters of galaxies
are even larger with diameters of 100 \Mpc\ and more.  Thus it was thought
that the homogeneity begins on supercluster scale.

Recent studies have shown that even on supercluster scale the distribution of
galaxies and clusters of galaxies may have some regularity.  First clear
indication for the presence of periodicity in the distribution of high-density
regions in the distribution of galaxies came from the deep pencil-beam survey
of redshifts of galaxies by Broadhurst \etal  (1990). High-density regions in
this survey form a regular pattern with a period 128 \Mpc. Nearest peaks in
the distribution of high-density regions in this survey coincide in position
and redshift with superclusters of galaxies (Bahcall 1991), thus one may think
that the distribution of superclusters may have some regularity. However,
as no clear periodicity was found in other directions, this result was
explained as a statistical anomaly (Kaiser and Peacock 1991).

Independent studies have shown that approximately on the same scale the
cluster correlation function has a weak secondary maximum (Kopylov \etal 
(1988), Mo \etal (1992), Einasto and Gramann (1993), Fetisova \etal (1993)).
This scale has been detected in the three-dimensional supercluster-void
network by Einasto \etal (1994), Einasto (1995a, 1995b).  The analysis of the
Las Campanas Redshift Survey has shown that the distribution of sheet-like and
filamentary structures has a preferred scale about 100~\Mpc\ (Doroshkevich
\etal 1996), also the two-dimensional power spectrum of galaxies of this
survey has a peak on the same scale (Landy \etal 1996). These results were
based on small number of objects or (with the exception of the study by
Einasto \etal 1994) on two-dimensional data, thus further studies are needed.

To investigate the distribution of matter we have used a new redshift
compilation of rich clusters of galaxies by Andernach, Tago and
Stengler-Larrea (see 1995). This is the largest and deepest
three-dimensional survey available presently.  Using this dataset we have
compiled a new catalogue of superclusters of galaxies (Einasto \etal  1997a),
calculated the nearest neighbour and void diameter distribution (Einasto \etal 
1997a), and determined the cluster correlation function and power spectrum
(Einasto \etal  1997b, 1997c, 1997d). Here I give a short summary of principal
results of these studies.

\section{The distribution of superclusters}

The compilation of Abell-ACO clusters of galaxies by Andernach, Tago and
Stengler-Larrea (1995) contains measured redshifts for 869 of the 1304
clusters with an estimated redshift up to $z=0.12$. For the present analysis
we used all rich clusters (richness class R$\ge 0$) in this compilation with
at least two galaxy redshifts measured. Cluster distances were determined from
redshifts or from the brightness of the clusters 10-th brightest galaxy,
using the photometric estimate of Peacock \& West (1993).  The new catalogue
contains 220 superclusters with at least two member clusters (supercluster
richness); 25 superclusters are very rich with at least 8 members,
approximately 25\% of all clusters are members of these very rich
superclusters.

The distribution of clusters in rich superclusters in supergalactic
coordinates is shown in Figure~1. We see that the population of clusters in
rich superclusters forms a fairly regular network.  The rectangular
distribution of clusters was noted by Tully \etal  (1992), who used the term
``chessboard universe'' to describe the structure.  Void diameter and nearest
neighbour tests indicate that the mean distance of rich superclusters across
voids is about $120 \pm 20$~\Mpc. Poor superclusters and isolated clusters lie
in the vicinity of rich superclusters, they also populate void walls but are
absent in central regions of voids defined by rich superclusters (Einasto
\etal  1994, 1997a).

{\small
\begin{figure}
\vspace{-1cm}  % amount of vertical space needed
{\epsfysize= 7 cm \epsfbox[-20 20 275 275]{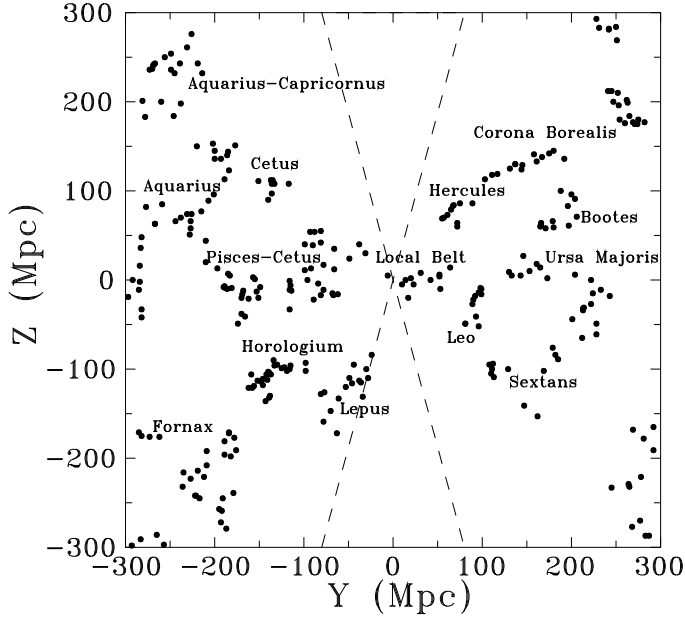}}
\caption{ The distribution of clusters of galaxies in
supergalactic coordinates, the sheet is taken in between
supergalactic $-100 \leq X <100$ {$h^{-1}$~{\rm Mpc}}.  Clusters
belonging to rich superclusters (containing at least 4 clusters)
are plotted with filled circles. Superclusters are identified by
common names (from constellations where they are located).
Dashed lines mark the zone of avoidance near the galactic
plane.}
\end{figure}
}

\section{ Cluster correlation function}

To quantify the regularity of the cluster distribution we have calculated the
correlation function and the power spectrum for clusters of galaxies.  The
correlation function describes the distribution of clusters in the real space,
the power spectrum in the Fourier space of density waves.  Analysis of various
geometric models has shown that if superclusters form a quasiregular lattice
with an almost constant step size then the cluster correlation function is
oscillating, it has alternate secondary maxima and minima, separated by half
the period of oscillations. The period of spatial oscillations of the
correlation function is equal to the step size of the distribution (Einasto
\etal  1997c).  The amplitude of the power spectrum at the wavelength
corresponding to that period is enhanced with respect to other wavelengths,
i.e. it is peaked. In contrast, if superclusters are located randomly in space
then the correlation function approaches zero level at large separations and
the power spectrum turns smoothly from the region with positive spectral index
on large wavelengths to a negative index on small wavelengths.

Correlation functions were calculated using the classical definition by
(Peebles 1980). The cosmic variance (error) of the correlation
function was determined following the method suggested by
Mo, Jing and B\"orner (1992). The cosmic variance depends on the number
of clusters in samples and does not depend on the bin size, the normalising
constant was determined from the scatter of realisations of various
N-body and geometric models (for details see Einasto \etal  1997c).

We have divided the whole sample of clusters into two populations, one
population in high-density regions (rich superclusters with at least 8
members), and the other in low-density regions (isolated clusters and clusters
in poor superclusters with a number of member clusters less than 8), Figure~2.
To suppress random errors the correlation function has been smoothed with a
Gaussian kernel of dispersion 15 \Mpc.  Our results show that the correlation
function of clusters in high-density regions has an oscillatory behaviour.
Maxima and minima alternate with a period of $\approx 120$ \Mpc.  The
$1\sigma$ error corridor is considerably smaller than the amplitude of
oscillations. A small overall decrease of the correlation function with
distance is due to inaccuracy of the selection function.  The population of
clusters in low-density regions has an uniform correlation function which
approaches zero on large scales.

{\small
\begin{figure}
\vspace{-1cm}  % amount of vertical space needed
{\epsfysize= 7 cm \epsfbox[10 40 250 285]{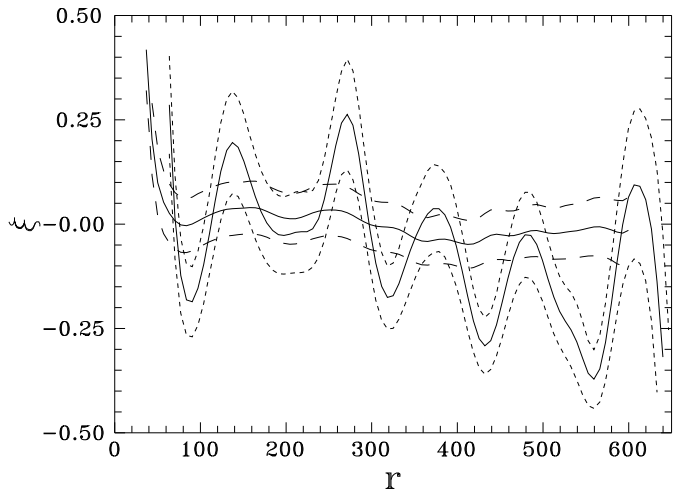}}
\caption{ 
The correlation function of clusters of galaxies. The solid line 
with dotted error corridor shows the correlation function of clusters
located in rich superclusters with at least 8 members, the solid line
with dashed error corridor gives the correlation function of isolated
clusters and clusters in poor superclusters.
}
\end{figure}
}

{\small
\begin{figure}
\vspace{-1cm}  % amount of vertical space needed
{\epsfysize= 7 cm \epsfbox[10 40 250 285]{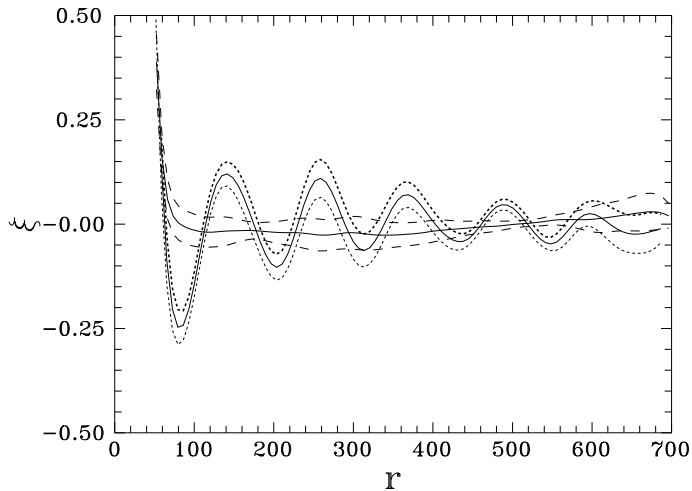}}
\caption{ 
Correlation functions of clusters in geometrical models. The solid curve with
dotted error corridor is the regular model with superclusters located
along regularly spaced rods, the solid curve with dashed error corridor
is the random supercluster model.
}
\end{figure}
}

The oscillatory behaviour of the correlation function is rather
surprising. The presence of the first secondary maximum of the cluster
correlation function was shown already by Kopylov \etal  (1988),
and confirmed by Mo \etal  (1992), Einasto and Gramann (1993) and
Fetisova \etal  (1993). Further maxima were  detected by Saar \etal 
(1995). 

To understand the oscillatory behaviour of the correlation function of
clusters of galaxies we have calculated the correlation function for randomly
and regularly located supercluster models.  In the second model superclusters
were located randomly along regularly spaced rods which form a rectangular
lattice of step size $120 \pm 20$~\Mpc\ as samples of real rich superclusters.
For each model we generated ten random realizations to calculate the error of
the correlation functions from the scatter of these realizations.

The correlation functions for our geometric supercluster models are given in
Figure~3.  As expected, the correlation function for the regular model is
oscillating as the correlation function of clusters in rich superclusters.
The correlation function of the random supercluster model has a large positive
correlation on small scales and zero mean correlation on large scales. Nearest
neighbour test, and pencil-beam and void analysis indicate that clusters in
poor superclusters with less than 8 members and isolated clusters form a
population, preferentially located in void walls between rich superclusters
but not filling the voids (Einasto \etal  1994, 1997a).  This analysis shows
that on large scales the correlation function characterises the regularity of
distribution of clusters of galaxies, differences between the random
supercluster model and actual distribution of poor superclusters in void walls
is irrelevant.

\section{Power spectrum of clusters of galaxies}

To calculate the power spectrum we have used the sample which 
contains clusters with measured redshifts only and lying in both
galactic hemispheres out to the distance covered by our cluster and
supercluster catalogues. The power spectrum was derived using two
different methods, a direct one where we calculate the distribution of
clusters in the wavenumber space, and an indirect method where we
first calculate the correlation function of clusters of galaxies and
then find the spectrum. In the latter case we make use of the fact
that the power spectrum and the correlation function are related by
the Fourier transform.

In both methods the main problem is the calculation of the selection
function of clusters of galaxies which corrects for incompleteness both at
low galactic latitude $b$ and at large distances $r$ from the observer.
The selection function can be represented by linear functions of $\sin b$
and $r$ (Einasto \etal 1997a, 1997b). Both methods to derive the power
spectrum yield similar results. However, parameters of the correlation
function are less sensitive to small inaccuracies of the selection
function, thus the indirect method yields more accurate results for the
power spectrum. 

To check the indirect method of calculation
of the power spectrum we have used simulated cluster samples having
similar selection effects as real ones. This check was performed for a
wide variety of models with different initial spectra. Our results
show that the true spectrum can be restored over the wavenumber
interval from $k\approx 0.03$~$h$~Mpc$^{-1}$ towards shorter waves
until $k\approx 0.3$~$h$~Mpc$^{-1}$.  

{\small
\begin{figure}
\vspace{-1cm}  % amount of vertical space needed
{\epsfysize= 7 cm \epsfbox[25 270 325 580]{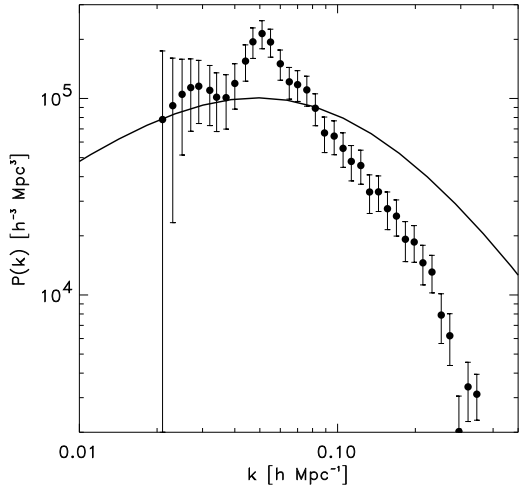}}
\caption{
The power spectrum for 869 clusters with measured
redshifts is plotted with solid circles. The spectrum is calculated
from the cluster correlation function via the Fourier transform.
Errors were determined from $2\sigma$ errors of the correlation
function found from the scatter of different simulations. 
The solid line is the standard
CDM ($h =0.5$, $\Omega = 1$) power spectrum enhanced by a bias
factor of $b=3$ over the four year COBE normalisation.
}
\end{figure}
}

The power spectrum for clusters of galaxies is shown in Figure~4.
On very large
scales the errors are large due to incomplete data. On moderate scales
we see one single well-defined peak at a wavenumber $k_0=0.052~h~{\rm
Mpc}^{-1}$.  Errors are small near the peak, and the relative
amplitude and position of the peak are determined quite
accurately. The wavelength of the peak is $\lambda_0=2\pi/k_0=120 \pm
15$~\Mpc. Near the peak, there is an excess in the amplitude of
the observed power spectrum over that of the CDM model (see below) by
a factor of 1.4. Within observational errors our power spectrum on
large scales is compatible with the Harrison-Zeldovich spectrum with
constant power index $n=1$, and on small scales with a spectrum of
constant negative power index $n=-1.8$.

Our calculation show also that spectra found for the sample of all clusters
and for the sample of clusters located in rich superclusters are very similar,
only the amplitude of the spectrum of clusters in rich superclusters is
higher. The sample of clusters in all supercluster richness classes is larger
than the rich supercluster sample and random errors of the spectrum are
smaller.  The sample of all clusters (with and without measured redshifts) is
still larger but in this case distance errors distort the correlation function
a bit more and the spectrum is less certain. For this reason we have used the
spectrum for all clusters with measured redshifts.

\section{ Correlation functions and spectra  in N-body models}

The power spectra and correlation functions are often used to compare the
distribution of matter in the Universe with theoretical predictions. 

We have calculated several models of structure formation using the standard PM
code with $128^3$ particles and $256^3$ cells.  Periodic initial conditions
were used in the computational volume of side-length $L=768$ \Mpc.  Four
initial spectra were used, corresponding to the standard CDM scenario with
$\Omega_{0}=1$ and Hubble constant $h=0.5$, a CDM model with cosmological
constant ($\Omega_{\Lambda}=0.7$, $\Omega_{0}=0.3$), a double power-law model
with spectral index $n=1$ on large scales, and index $n=-1.5$ on small scales,
and a transition at scale $\lambda_t=128$ \Mpc, and a similar double power-law
model with an extra peak near the maximum of the spectrum.
Clusters of galaxies were selected using a friend-of-friends algorithm for
test particles representing the clustering of dark matter particles. 
The cluster power spectrum of matter for the standard CDM model is
shown in Figure~4. 

Currently popular structure formation theories are based on the dynamics of a
Universe dominated by Cold Dark Matter.  Spectra of CDM-type models are rising
on long wavelengths $\lambda$ (small values of the wavenumber
$k=2\pi/\lambda$), and falling on small wavelengths (large values of $k$).
The transition between small and long wavelength regions in the CDM-spectra is
smooth. The distribution of superclusters in CDM-models is irregular (Frisch
\etal  1995). 

Spectra of double power-law models have a sudden transition between
small and long wavelength regions. Correlation functions of clusters of
galaxies of these models are oscillating. The amplitude of oscillations
is very large in the model with an extra peak near the maximum of the
spectrum. Rich superclusters in double power-law models form a moderately
regular network, the regularity is strong in the model with extra peak
in the spectrum.

The relative amplitude of the observed power spectrum above the standard
CDM-type model is not very large. Thus we may ask the question: within the
framework of the standard CDM-cosmogony, how frequently can we expect to find a
distribution of clusters which has a power spectrum and correlation function
similar to that observed?  To answer this question we determined the
correlation function and power spectrum for clusters in rich superclusters of
CDM-type models. In the spectral range of interest the power spectrum of the
standard CDM model is similar to the spectrum of a random supercluster model
(Einasto \etal  1997c). In both cases the correlation function of rich
superclusters in double conical volumes has randomly located peaks and
valleys.  We have generated 1000 realizations of the random supercluster
model, applied the selection effects as found in cluster distribution, and
determined the parameters of the cluster correlation function and power
spectrum.  To quantify oscillating properties of the correlation function we
measured the mean period and amplitude and their respective scatter.  We also
calculated the deviations for individual periods.  This test shows that
combination of parameters close to the observed values occurs in approximately
1~\% of cases, but the simultaneous concurrence of all parameters with
observations is a very rare event (of the order of one in million). Thus some
change in the initial spectrum of matter is necessary in order to explain the
observed correlation function and power spectrum for clusters of galaxies.

\section{ Conclusions}

Our study of the distribution of clusters of galaxies has lead us to
the following main conclusions.

\begin{itemize}

\item{} The distribution of high-density regions in the Universe (rich
  superclusters) is more regular than expected previously. Superclusters and
  voids form a cellular lattice or network with step size $120 \pm
  20$~\Mpc. The location of cells is rather regular.

\item{} The correlation function of clusters of galaxies has an oscillatory
  behaviour with regularly spaced secondary maxima and minima.  The period of
  oscillations, 120 \Mpc, is equal to the scale of the supercluster-void
  network.  The power spectrum of the cluster correlation function has a sharp
  peak on the respective wavelength.

\item{} Clusters of galaxies in CDM-type models of structure formation are
  located less regularly than real clusters. 

\item{} If the distribution of clusters of galaxies reflects the distribution
  of all matter then presently popular structure formation theories need
  revision.

\end{itemize}

ACKNOWLEDGEMENTS.  I thank my colleagues H. Andernach, M. Einasto, P. Frisch,
S. Gottl\"ober, V. M\"uller, V. Saar, A. A. Starobinsky, and E. Tago for
fruitful collaboration and for the permission to use our results prior to
publication. This work was partly supported by the grants from the
International Science Foundation, Estonian Science Foundation, German Science
Foundation and Russian Federation.

{}  

\end{document}